\documentclass[genTeX]{nrc1}
\usepackage[pctex32]{graphicx}

\journal{Can. J. Phys.}
\begin{document}

\title{The Standard Cosmological Model}
\author[Douglas Scott]{Douglas Scott}
\address{Department of Physics \& Astronomy, University of British Columbia,
Vancouver, B.C., V6T1Z1, Canada. \email{dscott@phas.ubc.ca}}

\shortauthor{Scott}

\maketitle
\begin{abstract}
The Standard Model of Particle Physics (SMPP) is an enormously successful
description of high energy physics, driving ever more precise measurements
to find `physics beyond the standard model', as well as providing motivation
for developing more fundamental ideas that might explain the values of its
parameters.  Simultaneously, a description of the entire 3-dimensional
structure of the
present-day Universe is being built up painstakingly.
Most of the structure is stochastic in nature, being merely the
result of the particular realisation of the `initial conditions' within our
observable Universe patch.  However, governing this structure is
the Standard Model of Cosmology (SMC), which appears to require only
about a dozen parameters.  Cosmologists are now determining the
values of these quantities with increasing precision in order to search
for `physics beyond the standard model', as well as trying to develop
an understanding of the more fundamental ideas which might explain the
values of its parameters.  Although it is natural to see analogies
between the two Standard Models, some intrinsic differences also exist, which
are discussed here.
Nevertheless, a truly fundamental theory will have to explain
{\it both} the SMPP and SMC, and this must include an appreciation of which
elements are deterministic and which are accidental.  Considering different
levels of stochasticity within cosmology may make it easier to accept that
physical parameters in general might have a non-deterministic aspect.
\end{abstract}
%
%%\begin{resume}
%%French version of abstract (supplied by CJP if necessary)
%%   \traduit
%%\end{resume}

\def\tablefootnote#1{% 
\hbox to \textwidth{\hss\vbox{\hsize\captionwidth\footnotesize#1}\hss}} 

\section{The Two Standard Models}
\label{sec:SMs}

Probably the most audacious endeavours in modern physical science are: (1) the 
attempt to understand the laws governing the whole of physical reality
down to the smallest imaginable scales;
and (2) the attempt to find a quantitative description
of the properties of the entire Universe on the largest scales.
Since the Universe is known to
be expanding and cooling, then these two quests become linked at the earliest
epochs, and hence fundamental physics and cosmology are necessarily
connected.

We do not yet have a complete theory to describe high energy physics, but
at below TeV energies our understanding is on extremely firm ground.
The combination of Quantum Chromodynamics with Electroweak Theory is known
as the Standard Model of Particle Physics (hereafter SMPP).  The SMPP was
solidified in the early 1970s and has been incredibly well tested since then
(see \cite{PDG} and earlier editions of the `Particle Data Book').

The SMPP contains a finite number of parameters,
which are unrelated, at least within the context of the theory itself.  One
imagines that a more complete theory of fundamental physics will explain the
relationships among these parameters.  The ultimate goal would be to
determine the values of the parameters from pure mathematics, once the
correct theory is discovered.  It may also be that some of the parameters
have a stochastic origin, where our Universe is one choice from among an
array of possible vacuum states.  This used to be called Anthropic reasoning
(see e.g.~\cite{BarTip,ReesCH,Carr}),
and received such little respect from many scientists that it became known
as `the A word', and would elicit groans at conferences.  But now string
theorists have renamed it `the Landscape' \cite{Suss}
and given it some theoretical basis.
Although these ideas may now have a little more mainstream credibility
(and are discussed in a later section), still
not everyone agrees that it is a worthy avenue of inquiry.\footnote{And it has
become known as `the other L word'.}

The number of parameters within the standard model varies slightly among
phenomenologists, depending on precisely how minimal the model under
consideration is, and, in particular, how the neutrinos are treated.
A popular counting exercise gives 19 parameters in the
minimal SMPP, plus 7 additional quantities to describe the neutrino sector.
This is shown in Table~\ref{tab:SMPP}.
There are 26 free parameters in this model;
if we were to develop the SMPP from scratch, then presumably we would
label the parameters as $A$, $B$, $C$, \ldots , $Z$.  Given this proliferation
of numbers, one expects that, for the sake of elegance, there must be a more
fundamental theory with far fewer parameters.

\begin{table} 
\captionwidth317pt 
\begin{center} 
\topcaption{The 26 Parameters of the Standard Model of Particle Physics.}
\label{tab:SMPP} 
\begin{tabular}{lcccccc}
\hline* 
\phantom{1}6 quark masses: & $m_u$ & $m_d$ & $m_s$ & $m_c$ & $m_t$ & $m_b$\\
\phantom{1}4 quark mixing angles: & $\theta_{12}$ & $\theta_{23}$ &
 $\theta_{13}$ & $\delta$ & & \\
% & \multicolumn{4}{c}{CKM matrix} & & \\
\phantom{1}6 lepton masses: & $m_e$ & $m_\mu$ & $m_\tau$ & $m_{\nu_e}$ &
 $m_{\nu_\mu}$ & $m_{\nu_\tau}$ \\
\phantom{1}4 lepton mixing angles: & $\theta^\prime_{12}$ &
 $\theta^\prime_{23}$ & $\theta^\prime_{13}$ & $\delta^\prime$ & & \\
\phantom{1}3 electroweak parameters: & $\alpha$ & $G_{\rm F}$ & $M_Z$ & & & \\
\phantom{1}1 Higgs mass: & $m_{\rm H}$ & & & & \\
\phantom{1}1 strong CP violating phase: & $\bar\theta$ & & & & \\
\phantom{1}1 QCD coupling constant: & $\alpha_{\rm S}(M_Z)$ & & & & \\
\hline 
26 total parameters & &&& \\
\hline* 
\end{tabular} 
\end{center} 
\end{table} 

As is well known, the SMPP
has been astonishingly successful, so much so that,
for the last 3 decades, the emphasis has been on trying to find inadequacies
in it -- i.e.~searching for `physics beyond the standard model'.
However, apart from theoretical ideas (some of them admittedly quite
appealing), there are still {\it no\/} convincing pieces of evidence for
physics beyond the SMPP.

On the other hand, we {\it know\/} that there has to be new physics, beyond
the SMPP, due to what we have learned about the properties of the large-scale 
Universe -- particularly cosmological evidence for dark matter, dark energy
and inflation.

Cosmology grew from being an arm-chair activity carried out in people's
spare time, to being a dignified scientific pursuit, only in the 1960s.
Originally the models were entirely baryonic and involved simple ad hoc
initial conditions.  In many ways the basic picture has remained the same
since then -- nearly scale invariant and adiabatic initial conditions, in an
almost isotropic and homogeneous Friedmann-Robertson-Walker solution to
Einstein's Field Equations.  However, Cold Dark Matter was added to the
paradigm in the 1980s (e.g.~\cite{Peebles82,BFPR}), leading to the
`Standard CDM' picture in which $\Omega_{\rm M}=1$.  By the end of
the 1980s the addition of a cosmological constant $\Lambda$
was known to give better fits to the available data
(e.g.~\cite{Peebles84,Vittorio85,Efstathiou90}).

The {\sl COBE\/} satellite detection of large-scale Cosmic Microwave Background
(CMB) anisotropies in 1992 \cite{Smoot92}
brought an end to many wilder proposals which had been floated in the era
of continually improving CMB upper limits (see \cite{Romans} for a discussion).
It became clear that the CMB normalization, together
with galaxy clustering data, pointed to the `$\Lambda$CDM' variant of
the CDM paradigm (\cite{EBW,Kofman93}), despite the reluctance of many
theorists to let the elegance of Standard CDM slip away
(e.g.~\cite{Resuscitated}).  The cosmological constant became an accepted
part of the model by the mid-to-late 1990s, following the results from distant
supernova surveys and degree-scale CMB experiments.  Soon the concept of
$\Lambda$ was generalised
to that of Dark Energy.  As the CMB anisotropy measurements grew
increasingly precise, it became clear that (at least in principle)
several parameters could be
measured which would constrain the inflaton potential.  But to do this
carefully, one had to take into account other astrophysical effects on the
CMB anisotropies, particularly anisotropy suppression in the period since the
Universe became reionized -- hence another parameter needed to be added.

We have thus ended up with a Standard Model of Cosmology (hereafter SMC),
which is based on ideas as old as the SMPP, but which solidified only about
a decade ago.  Determining the precise date when the SMC was in place is a
little murky (to say the least).
The late 1980s and early 1990s were a time of increasing
tension among different pieces of observational data, which (at least
in hindsight) was because the SMC was about to fall into place.  There were
also a few false leads, such as the early supernova results apparently
suggesting {\it deceleration}, increased interest in models with a
significant hot dark matter (i.e.~high $\Omega_\nu$) component, and
arguments over the naturalness of open inflationary models.
But despite all of this, the SMC was clearly in place by 1995
\cite{KraTur,OstSte}.

The number of parameters required to describe this model varies to some
extent depending on the tastes of individual cosmologists.  However, a
typical count gives the number of required parameters as 12, which are
listed in Table~\ref{tab:SMC}.  This is not a complete set of possible
parameters, but there is currently no evidence that we need any more.
If we were to develop the SMC from scratch, then presumably we would choose
a simpler set of symbols, for example:
$A, E, H, I, K, L, M, N, O, P, U, W$.\footnote{The Hawaiian alphabet.}
The parameters are also not all on an equal footing.
For some of them, there is no indication at the moment that they differ from
their default values (e.g.~$\sum\Omega_i=1$ or $n^\prime=0$), and hence the
final SMC may actually have {\it fewer\/} parameters.

\begin{table} 
\captionwidth317pt 
\begin{center} 
\topcaption{The 12 Parameters of the Standard Model of Cosmology.}
\label{tab:SMC} 
\begin{tabular}{lcccc} 
\hline* 
\phantom{1}1 temperature: & $T_0$ &&& \\
\phantom{1}1 timescale: & $H_0$ &&& \\
\phantom{1}4 densities: & $\Omega_{\Lambda}$ & $\Omega_{\rm CDM}$ &
 $\Omega_{\rm B}$ & $\Omega_\nu$\\
\phantom{1}1 pressure: & $w\equiv p/\rho$ &&& \\
\phantom{1}1 mean free path: & $\tau_{\rm reion}$ &&& \\
\phantom{1}4 fluctuation
descriptors: & $A$ & $n$ & $n^\prime\equiv dn/d\ln{k}$ & $r\equiv T/S$ \\
\hline 
12 total parameters & &&& \\
\hline* 
\end{tabular} 
\end{center} 
%\tablefootnote{The 12 parameters of the SMC.}
\end{table} 

There are several assumptions that underlie the SMC.  We certainly assume
that physics is the same everywhere in the observable Universe (but
see Section~\ref{sec:stoch}), and that
General Relativity fully describes gravity on large scales.  The SMC also
relies on the hot Big Bang picture being correct, and that something akin to
inflation created the density perturbations.  The astonishing thing about
modern cosmology is that most of these assumptions are testable (or at least
falsifiable), and that for the reality in which we find ourselves living there
are ways of determining the values of the quantities that describe the
nature of the entire observable Universe.

\section{The Miracle of the CMB Sky}
\label{sec:cmb}

Many different observable quantities can be used to constrain
the cosmological parameters.  Traditionally these have involved trying to
estimate distances of very distant objects (which is hard), estimating
masses of large amounts of matter (which requires the distance), measuring
the clustering of galaxies (which is related in a complicated way to the
clustering of mass), and determining primordial abundances (which is
fraught with systematic effects).  While each of these approaches have been
useful, they {\it all\/} rely on using tracers that are well into the
non-linear regime, i.e.~objects with density contrasts $\delta\gg1$.

\begin{figure}
\begin{center}
\topcaption{Theoretical CMB anisotropy power spectrum.  This is the expected
variance per logarithmic interval in $\ell$ for the SMC (together with an
arbitrarily normalised tensor component).  The physics of the CMB anisotropies
splits into 4 regions, separated approximately into decades of $\ell$ (see
\cite{ScoSmo} for more details).}
\includegraphics[height=10truecm]{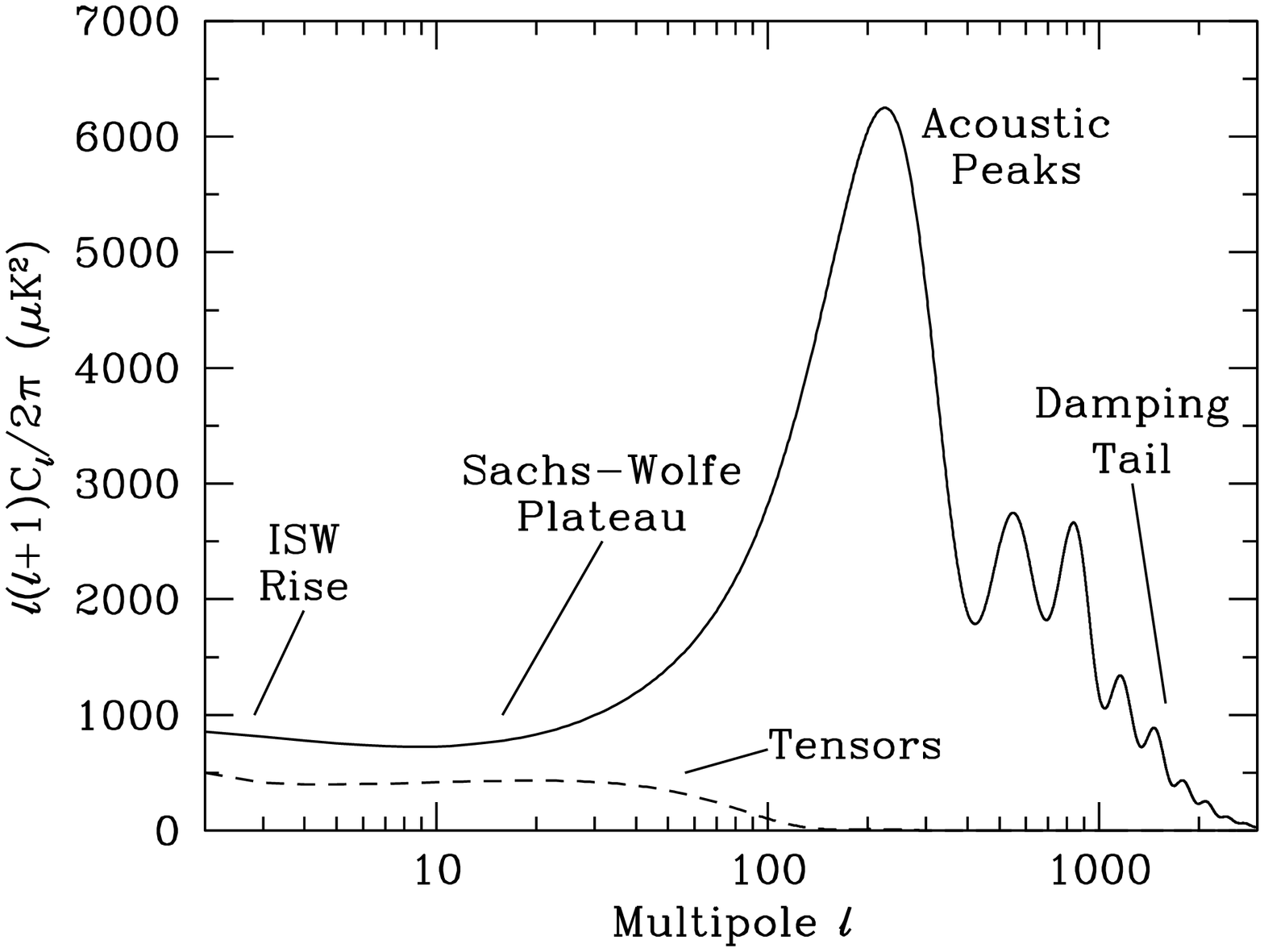}
\label{fig:CMBtheory}
\end{center}
\end{figure}

Fortunately the Universe has given us another opportunity to learn about
its large scale properties -- anisotropies in the Cosmic Microwave Background
(see e.g.~\cite{WSS,HuSugSil,HuDod,ScoSmo} for reviews).
These are essentially
a projection of the 3-dimensional structure at early times
(redshift $z\sim1000$, or $t\sim300{,}000$ years), when densities were still
very much in the linear regime, $\delta\sim10^{-5}$.

Apart form the dipole (see later), the CMB is extremely isotropic.
This `horizon problem'
is often taken to be evidence that the initial conditions were acausal, or
that something happened to make them appear so.  But more important information
comes from the fact that the CMB actually contains minute anisotropies,
first detected by the {\sl COBE\/} satellite, and confirmed and extended
by about 20 separate experiments in the last 13 years.  The {\it particular\/}
anisotropies that we observe are thought to be stochastic, in the sense that
different volumes of the Universe will have different CMB skies.  But
the power spectrum of anisotropies, from which our sky's temperatures are
drawn, is considered as a fundamental quantity, which depends on the
underlying cosmological model.

If the density perturbations are Gaussian, then the power spectrum describes
{\it all\/} of the statistical information.  Non-Gaussianities may
certainly exist and are worthy of study, but they are weak in all
inflationary-type models, and hence the power spectrum contains the vast
majority of the information and has therefore become the main focus of all
modern CMB experiments.

Figure~\ref{fig:CMBtheory}
shows a calculation of the power spectrum of CMB anisotropies,
where $C_\ell\equiv \left| a_{\ell m} \right |^2$ and the $a_{\ell m}$s are
the amplitudes of the spherical harmonic expansion of the temperature
field.  The structure of this power spectrum splits into 4 regions,
each of which conveniently takes up about a decade in $\ell$.  The peaks arise
from acoustic modes in the coupled photon-baryon fluid (think of these as
a snapshot of the amplitudes of standing waves at the last scattering
epoch), which damp at higher $\ell$s because of incomplete coupling.  
At larger angular scales we have the Sachs-Wolfe effect, which is an
imprint of the initial conditions, seen because of gravitational redshifts
through potentials,
together with the `Integrated Sachs-Wolfe' effect for the very lowest
$\ell$s (from the line of sight integral term).

CMB anisotropies can be determined with exquisite precision for specific
cosmological models \cite{HSSW,SSWZ}.  This is because the physics
generating the anisotropies is well understood, requiring just linear
perturbation theory for gravity-driven acoustic oscillations,
together with simple Thomson scattering of the CMB photons off free
electrons.

\begin{figure}
\begin{center}
\topcaption{CMB band-powers from the 
{\sl WMAP\/} \cite{Hinshaw03}, BOOMERANG \cite{Jones05},
VSA \cite{Dickinson04}, CBI \cite{Readhead04} and ACBAR \cite{Kuo04}
experiments.  Some of the higher
and lower $\ell$ band-powers with the poorest error bars have been omitted.
There are also some other experiments with data of comparable quality.}
\includegraphics[height=10truecm]{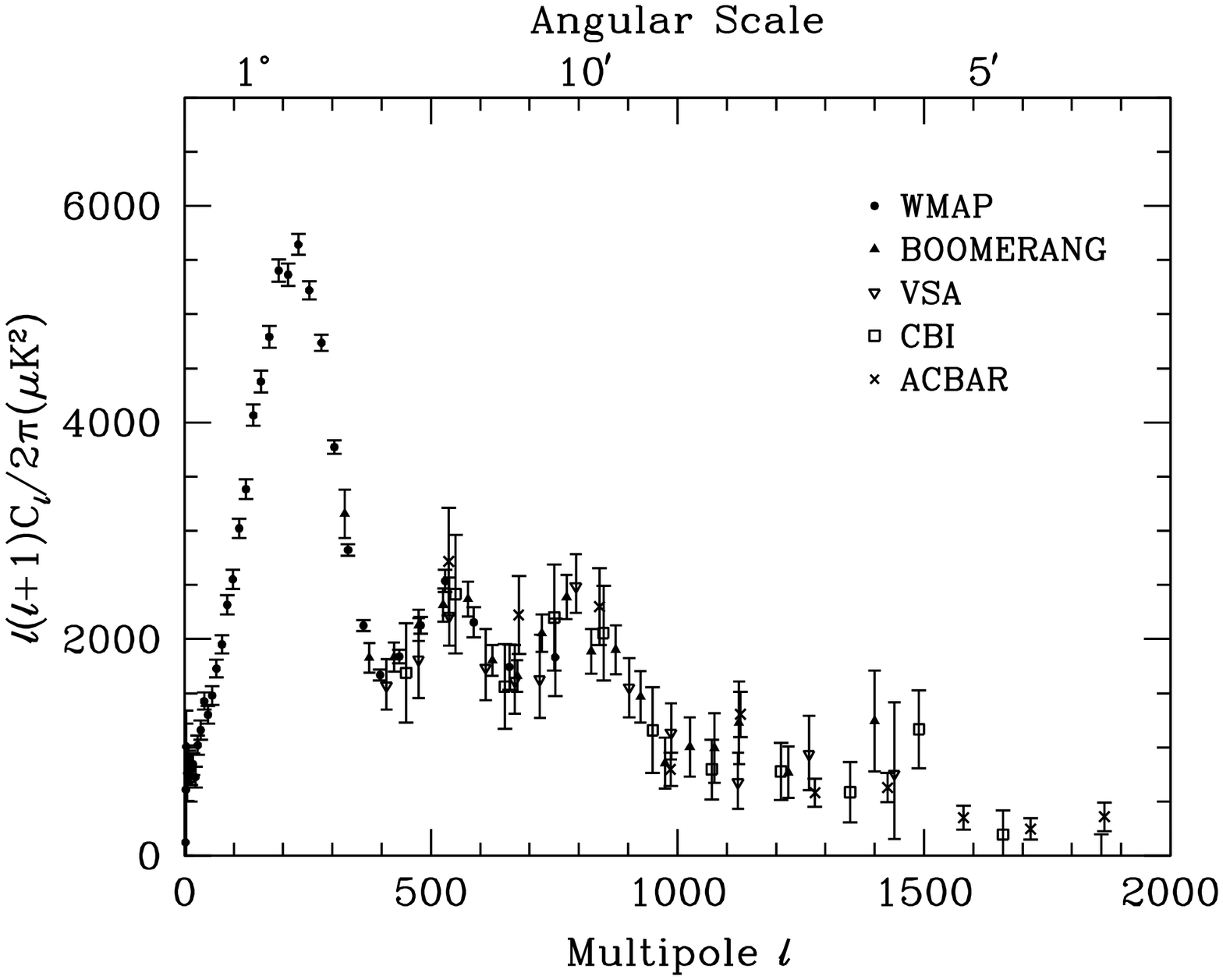}
\label{fig:CMBdata}
\end{center}
\end{figure}

Experimental results are usually quantified in terms of `band-powers'
on the anisotropy power spectrum.  A compilation of some of the highest
quality data-sets is shown in Figure~\ref{fig:CMBdata}
(note, there is a linear $x$-axis here, unlike in Figure~\ref{fig:CMBtheory}).

Clearly one does not need a model to be plotted to guide the eye on this
figure.  The first three
acoustic peaks are now convincingly detected.  So the general paradigm, of
a hot early Universe, containing roughly scale-invariant and adiabatic
initial perturbations over a wide range of scales, is strongly supported.
It is easy to see how considerable cosmological information is encoded
in the rich structure of the power spectrum.  In fact a 6 parameter model
space is very tightly constrained by the current CMB
data \cite{Spergel03,MacTavish05}, and an extended set of parameters can
be constrained using other astrophysical measurements.

The most precise quantity measured from CMB anisotropies is the angular
scale of the acoustic peak structure in the power spectrum.  This is often
parameterised by $\theta$ \cite{Kosowsky02}, defined to be 100 times the ratio
of the sound horizon to the angular diameter distance at last scattering.
It turns out that, defined this way, the value is remarkably close to unity,
with the best value somewhere around $1.045\pm0.004$ \cite{MacTavish05}.
This quantity is
related to the underlying cosmological parameters -- for example, it depends
strongly on the overall curvature of space.

The CMB does not of course measure everything.  Several degeneracies in fitting
the SMC parameters are easily broken by including other data, e.g.~from
supernovae, galaxy clustering
or direct estimates of $\Omega_{\rm M}$ or
$H_0$.  These data can be used as priors when fitting parameters to CMB data.
Hence, one must be careful when using particular results, to consider what
priors were used, as well as whether restricted parameter spaces or ranges of
parameters were considered.

The CMB is about 10\% polarized, since Thomson scattering of an anisotropic
radiation field generates linear polarization (see e.g.~\cite{HuW,Zal03}).
Measurements of the polarization
power spectrum and the cross-power spectrum of polarization with
temperature are also
now being measured.  So far such measurements have only served to confirm the
basic paradigm of the SMC (but see~\ref{sec:reion}), although it is
expected that more precise polarization measurements will help break
degeneracies and place quite different constraints on parameters.

Another particularly promising cosmological measurement technique
now being developed is weak gravitational lensing (also called `cosmic
shear', see e.g.~\cite{LVWYM}).  This approach to
measuring cosmological fluctuations shares some of the benefits (and much of
the mathematics) of CMB anisotropies.  Like CMB polarization, the results
so far have confirmed the SMC picture, but have not set very strong
additional constraints.  That will likely change with the more ambitious
lensing projects now underway.

\section{The Values of the Parameters in the SMC}
\label{sec:params}

Now that the SMC has been outlined and some of its observational probes
described, let us examine each of the parameters in turn.

\subsection{The temperature}
\label{sec:T0}

The CMB was discovered by Penzias \& Wilson
in 1965 after a convoluted history of false starts and neglected theoretical
predictions.  It was eventually found to have a very accurately blackbody
spectrum, with deviations severely constrained over 3 decades in wavelength
(see \cite{SmootScott02} and references therein).
The CMB temperature is measured to be
\begin{equation}
T_0= 2.725 \pm 0.001\,{\rm K},
\end{equation}
where the error bar represents $1\sigma$, but is dominated by systematics
\cite{Mather99}.
The fact that this is such a good blackbody is one of the strongest pieces
of evidence for the hot Big Bang picture.  It is extremely difficult, if not
impossible, to contrive to make such a smooth thermal spectrum through some
local process, but it is easy to achieve in a hot early period of the
Universe, since the thermal equilibrium timescale is naturally very much
shorter than the expansion time.

This temperature corresponds to a photon number density of
$n_\gamma\simeq411\,{\rm cm}^{-3}$ and an energy density of
$\rho_\gamma\simeq0.260\,{\rm eV}\,{\rm cm}^{-3}$ for a thermal spectrum.
One can also derive other quantities, such that the intensity peak is
$I_\nu\simeq385\,{\rm MJy}\,{\rm sr}^{-1}$ at
$\nu\simeq160.2\,{\rm GHz}$, that the r.m.s. magnetic field in the CMB
corresponds to $B_\gamma\simeq3.24\,\mu{\rm G}$, etc.

Another quantity we can determine from observations of the CMB is the
size of the dipole, i.e.~the amplitude of a $\cos\theta$
function fit
to the whole sky once the overall monopole amplitude $T_0$ has been removed.
This tells us our `local' velocity through the sea of CMB photons.
This velocity is generated by the gravity field of lumps
on the scale of the local Supercluster, and hence varies between different
observers in the Universe.  In other words this is a quantity that is
established by the particular realisation of the
observable Universe around us.

The observed dipole implies that the Solar System is moving at
$v\simeq370\,{\rm km}\,{\rm s}^{-1}$ towards a particular direction
(just above the equatorial plane at about $11^{\rm h}$), or, correcting for
estimated velocity vectors, one can find a speed for the Local Group of
galaxies corresponding to about $630\,{\rm km}\,{\rm s}^{-1}$ \cite{Fixsen96}.

The value of $T_\gamma$ changes with cosmological epoch so that
$T_\gamma(z)=T_0 (1+z)$.
There {\it is\/} some evidence that its value was higher in the past,
consistent with cooling through expansion (e.g.~\cite{Molaro02}).  But it would
take a Hubble time to observe a substantial change in its value (and hence
extraordinary precision to detect a change over a human lifetime).  Therefore,
we should regard $T_0$ as a fundamental constant, which fixes the energy
density of radiation today.

Although Alpher, Gamow \& Herman \cite{AlpHer} made an order of magnitude
prediction of the CMB temperature (based on assuming that while nuclei
were undergoing nuclear reactions in the early Universe they were in
thermal equilibrium with photons),
there is no ab initio calculation for the CMB temperature today.\footnote{For
some numerological ideas
see {\tt http://www.astro.ubc.ca/people/scott/whochosetemp.html}}
Hence $T_0$ should be regarded as an empirical parameter of the SMC.
Its value may be related to the details of: (a) reheating at the end of
inflation; (b) the SMPP and its extensions,
defining the particles which annihilate in the early Universe to give
most of today's CMB photons; and (c) the particular time that we happen to
exist and make measurements.

\subsection{The timescale}
\label{sec:H0}

The derivative of the cosmological scale factor measured today,
$H_0\equiv \left.({\dot a}/a)\right|_{t=t_0}$, is referred to as the Hubble
constant.  Locally it provides the calibration of the Hubble expansion
law, $v=H_0 r$, and hence is measured in units of
${\rm km}\,{\rm s}^{-1}{\rm Mpc}^{-1}$ (and sometimes defined as
$h\equiv H_0/100$ in these units).  The best direct estimate comes from
the Hubble Space Telescope Key Project \cite{Freedman01}.  This is
often used as a prior for the determination of parameters from CMB
anisotropies.  The recent multi-parameter fit by the BOOMERANG team gives
$H_0=75.8^{{+}5.6}_{{-}5.1}$, and with the addition of further constraints
from large-scale structure they find \cite{MacTavish05}
\begin{equation}
H_0=69.6\pm2.4\,{\rm km}\,{\rm s}^{-1}{\rm Mpc}^{-1}.
\end{equation}

The Hubble constant evolves according to $H(z)=H_0\sqrt{\Omega_{\rm M}(1+z)^3
+ \Omega_\Lambda}$ for flat cosmologies (i.e.~$\Omega_{\rm M}+\Omega_\Lambda=1$,
and for recent epochs where we can ignore radiation).  So we have to wait
of order the Hubble time to observe it change dramatically.  In fact, in a
purely cosmological constant dominated model, the value approaches a
constant as the Universe approaches the de Sitter limit:
$H(z)\to\sqrt{\Omega_\Lambda}H_0$ as $z\to-1$, giving a final value a little
below 60 for the currently preferred SMC parameters.
\footnote{So one could say that Sandage was right, he was just
billions of years ahead of his time!}

But even although $H_0$ is varying with time, we can regard it as a
constant for all practical purposes.  Since the Hubble constant is a rate,
then its reciprocal defines a timescale.  For the parameter values of the
current
SMC it turns out (through a coincidence, because of the combined effects of
deceleration and acceleration) that the age of the Universe is
$t_0\simeq 1/H_0$ to within 10\%.

Once one has a specific set of parameters for the SMC (actually just
the $\Omega$s and $H_0$), one can derive the present-day age of the Universe
precisely.  Hence $t_0$ is a derived quantity, with current
estimates in the range 13--$14\,$Gyr.

Since $H_0$ changes with time, then its value would have been different if
we happened to live, say, 5 billion years in the past or the future.  But
in a $\Lambda$-dominated universe a final value of $H_0$ is eventually
reached.  However,
even this value, $\sqrt{\Lambda/3}$, is not independent of the other
parameters, since it just depends on the cosmological constant.

\subsection{The densities}
\label{sec:omegas}

The amounts of each of the components that make up the Universe are
usually defined in terms of their contributions to the critical mass (or
energy) density.
Here $\rho_{\rm crit}\equiv3H_0^2/8 \pi G$ is the density required to make the
Universe spatially flat.
The total matter density $\Omega_{\rm M}\equiv \Omega_{\rm CDM}+\Omega_{\rm B}$
can be estimated fairly directly using various dynamical methods, which
typically give values of 0.2--0.3.  The baryonic contribution (which
traditionally includes electrons) can be separately
constrained by Big Bang Nucleosynthesis \cite{FieSar}

The CMB places a strong constraint on the overall curvature of space,
essentially through the $\theta$ parameter, i.e.~the $\ell$
scaling of the acoustic peaks.  This means that
$\Omega_{\rm tot}\equiv \sum\Omega_i$
(which is basically $\Omega_{\rm M}+\Omega_\Lambda$) is
strongly constrained to be close to unity.\footnote{And so the non-Euclidean
isotropic spaces which used to form an important part of all basic
cosmology courses can now be relegated to a footnote.}
Other than this geometric effect, the CMB anisotropies depend on the
physical densities of baryons and CDM, i.e.~$\rho_{\rm B}
\propto \Omega_{\rm B}h^2$ and $\rho_{\rm CDM}
\propto \Omega_{\rm CDM}h^2$.

Useful additional constraints, with different parameter dependencies,
come by adding data from the clustering of galaxies, e.g.~from the
Sloan Digital Sky Survey \cite{Pope04} or 2 Degree Field Galaxy Redshift
Survey \cite{Cole05}.  These large-scale structure (LSS) data depend mainly on
the combinations $\Omega_{\rm M}h$ and $\Omega_{\rm B}/\Omega_{\rm M}$,
as well as the normalization (called `bias') of the
galaxy clustering relative to that of the dark matter.

The newest BOOMERANG results, in combination with {\sl WMAP\/} and other
CMB experiments, and using some LSS information, yield
\cite{MacTavish05}
\begin{equation}
\Omega_{\rm B} h^2=0.0227\pm0.0008
\end{equation}
\begin{equation}
{\rm and}\ \Omega_{\rm CDM} h^2=0.120\pm0.005.
\end{equation}
The precise values depend on how much freedom is allowed on the shape of
the primordial power spectrum and which LSS data are
used.

The constraint on the cosmological constant (or Dark Energy)
using CMB data plus LSS
data is
\begin{equation}
\Omega_\Lambda=0.67\pm0.05,
\end{equation}
and tighter constraints are possible if one restricts the Universe to
be flat.  The best current limits show that $\Omega_{\rm tot}$ is within
a few per cent of unity.

A minor (but not negligible)
contribution to the overall energy density comes from massive
neutrinos.  They have the effect of partially erasing fluctuations on
small scales.  This process depends on the sum of the neutrino masses, assuming
they are Fermi-Dirac distributions with zero chemical potential and
the temperature expected in the Big Bang picture,
i.e.~$T_\nu\simeq(4/11)^{1/3}T_\gamma$.   The derived constraint on
$\Omega_\nu$ depends sensitively on the data which are used for smaller scale
clustering, as well as the freedom allowed on the primordial spectrum and how
one treats galaxy bias.   A fairly robust limit is
\begin{equation}
\Omega_\nu<0.02,
\end{equation}
corresponding to $\sum m_\nu<1\,$eV, although tighter constraints have also
been claimed (see e.g.~\cite{PisSer}).

As a fraction of the critical density, we find (using $T_0$) that
$\Omega_\gamma h^2=2.471\times10^{-5}$ for the photon contribution.
This value is small enough that it is
negligible in today's Universe, although one needs to
properly account for it when considering evolution at early times.
Apart from components we have already mentioned,
the only other contribution to the energy density is from gravity waves,
but this makes a fraction if $\rho_{\rm crit}$ which is several orders of
magnitude lower still.

The $\Omega$s change with time, and hence they cannot be regarded as
`fundamental' (although, like $T_0$ and $H_0$ they vary negligibly over a
human lifetime).  Relative to each other, the matter contributions do not
evolve, but the Dark Energy contribution is growing with time (largely because
the matter density is dropping, while Dark Energy is roughly constant per
unit volume).

We can also consider the underlying physics which determines
each of the individual density parameters:
$\Omega_\Lambda$ today is determined if we already know the
vacuum (or Dark Energy) density $\Lambda$ plus $H_0$; $\Omega_\nu$ depends
on the neutrino masses, as well as $T_0$ and $H_0$; $\Omega_{\rm CDM}$ is
presumably set by the physics of the dark matter particles (a mass and a
cross-section); and $\Omega_{\rm B}$ is set by the physics of baryogenesis
(for a given $T_0$).

\subsection{The pressure}
\label{sec:w}

Pure vacuum energy (a.k.a.~the cosmological constant) is fixed per unit volume,
and hence evolves as $\rho\propto (1+z)^0$, while `curvature' scales as
$(1+z)^2$.  Vacuum has pressure $p=-\rho$, while curvature behaves like a
fluid with $p=-(1/3)\rho$.  Hence the generalisation of $\Lambda$ is to a
`Dark Energy' (see e.g.~\cite{Carroll01})
with an equation of state parameter $w\equiv p/\rho$ in
the range $-1<w<-1/3$.  For empirical purposes, this range should be
extended to $w<-1$, although there are certainly theoretical difficulties
with such `phantom energy'.

The CMB alone is not very sensitive to the Dark Energy content, apart
from its contribution to $\Omega_{\rm tot}$.  But
additional constraints, particularly from supernovae data
(e.g.~\cite{Riess04,Astier05}) give upper limits around
\begin{equation}
w< -0.7,
\end{equation}
(see e.g.~\cite{MacTavish05,Seljak05}).  This is entirely consistent with
a pure cosmological constant.
If it were discovered that $w\neq -1$, then that would be extremely exciting,
since from its behaviour we could hope to understand what the Dark Energy
actually {\it is}.  Although there are plenty of
alternative names that have been dreamed up by theorists,
there are no well motivated theoretical models for Dark Energy.  Hence there
is no calculation that can tell us how small to expect $w+1$ to be.

\subsection{The mean free path}
\label{sec:reion}

We know (from the spectra of distant quasars) that the Universe reionized at
some redshift higher than 6 (see \cite{BarLoe} for a review).
Presumably this was achieved by the ionizing
photons generated in the earliest stars.  Understanding the `End of the
Cosmic Dark Ages' is one of the most exciting issues in modern
astrophysics.  Fortunately the CMB can help, since the extra Thomson
scattering from $z=0$ out to $z=z_{\rm reion}$ partially erases the
anisotropies.  This effect is hard to separate from an overall change
in amplitude and/or slope, {\it except\/} that it also generates a
large-angle polarization pattern.

The CMB fluctuations are expected to be weakly polarized
and with a pattern that is correlated with the anisotropies.  The {\sl WMAP\/}
experiment was able to measure the temperature-polarization correlations on
the largest angular scales (above $10^\circ$),
where a signal was detected that appeared to be
consistent with that expected from reionization, except somewhat
stronger than would have been guessed \cite{Kogut03}.
The estimated value for the Thomson optical depth
($\int n_e \sigma_{\rm T} dr$) is
\begin{equation}
\tau=0.17\pm0.04.
\end{equation}
This result can be recast into a mean free path or a redshift, given
the other background cosmology parameters, e.g.~it
corresponds to  $9<z_{\rm reion}<30$.  The error bar is still quite
large, however, and so somewhat  different values can be obtained using
other choices of data, priors, etc.

Clearly $\tau_{\rm reion}$ is calculable in principle, since it depends on
small scale density perturbations going non-linear and turning into stars.
However, this requires knowing details of the inflationary power
spectrum at smaller scales than are otherwise probed, plus understanding
complexities of `gastrophysics', which are currently well beyond the
capabilities of numerical simulations.  So for the foreseeable future
$\tau_{\rm reion}$ should be regarded as an empirical parameter which has to
be taken into account when fitting the CMB (including polarization)
anisotropies.  One may regard $\tau_{\rm reion}$ as non-fundamental, but
in the sense that its value is an outcome of the (unknown) underlying theory,
it is really no different than, say, $T_0$.

\subsection{The fluctuation descriptors}
\label{sec:flucts}

The adiabatic nature of the CMB anisotropies at angular scales above those
corresponding to the Hubble length at last scattering suggests there were
`synchronized initial conditions' at apparently acausal scales.  More direct
evidence for this comes from the temperature-polarization anti-correlation
seen by {\sl WMAP\/} above degree scales \cite{Kogut03}, which gives
a clearer indication for an early period of cosmic acceleration than the
observed near flatness, Gaussianity and scale-invariance.  Although there is
undoubtedly convincing evidence for
{\it something\/} here,\footnote{A defensible statement
is `Something like inflation is something like proven.'} we need more
direct probes of the physics of inflation in order to understand whether
it really happened (see \cite{LytRio} for a review).

There are 4 basic observables which can help probe the initial perturbations
-- 2 amplitudes, a slope and a curvature.  The 2 amplitudes are for scalar
modes (density perturbations) and tensor modes (gravity waves).  They are
usually parameterised with $A$ for the variance of the scalars at a fixed
wavenumber (e.g.~$0.05\,{\rm Mpc}^{-1}$), together with the tensor to scalar
ratio $r$ at even larger scales.  Then there is the slope of
the scalar power spectrum,
i.e.~$P(k)\propto k^n$.  This spectral index $n$ can also have a `running',
i.e.~one can potentially measure the curvature through
$n^\prime\equiv dn/d\ln k$.  Higher derivatives seem destined to be
unobservable, as does the tensor spectral index (although in most models it is
fixed by the value of $r$ in any case).

The results from the {\sl WMAP\/} experiment first year data are
\cite{Spergel03}
\begin{equation}
A=2.7(\pm0.3)\times10^{-9},
\end{equation}
\begin{equation}
n=0.97\pm0.03
\end{equation}
\begin{equation}
{\rm and}\ r<0.53.
\end{equation}
Again, slightly different constraints can be obtained by combining with
different data-sets and by restricting the parameter space being
considered (particularly whether one allows for $n^\prime\neq 0$).

For the running of the spectral index there appears to be weak evidence
that the slope changes from blue to red as one moves from small $k$ to
large $k$, with estimates around
\begin{equation}
n^\prime = -0.06\pm0.03.
\end{equation}
However, the precise value depends on the data-sets and priors used
\cite{Seljak05,MacTavish05}.  The result is currently $2\sigma$ at best, and
hence there is no strong evidence for non-zero running.
However, a convincing detection would
be quite constraining for inflationary models \cite{Peiris03}.

In slow roll inflation $A\propto V^3/(V^\prime)^2$, where $V(\phi)$
is the inflaton potential.  The slope $n$ also involves $V^{\prime\prime}$,
while $n^\prime$, etc. involves higher derivatives.  The nice thing is that
the tensor amplitude is proportional to $V$ itself.  Hence if one could
measure the gravity wave contribution to the CMB anisotropies (see
Figure~\ref{fig:CMBtheory}), then one would have an estimate of the energy
scale of inflation.  And if one could measure $A$ and $n$ precisely, then
one would be constraining a Taylor expansion of the potential.
But extracting a low amplitude tensor component from the temperature
$C_\ell$s is clearly difficult.

One of the most exciting prospects for CMB polarization measurements is
that they possibility provides a clean way of extracting the tensor
signal.  The polarization `B-mode' (a geometrical component of the pattern
which contains a curl)
cannot be sourced by scalars, since they have no handedness,
while primordial gravity waves {\it do\/} generate B-modes.  Hence there is
a great quest underway to improve the sensitivity of polarization experiments
to try to detect this signature, if it is there.  This will not
be easy, but the pay-off makes it worth the effort.

Presumably these 4 inflationary parameters come from some more fundamental
theory.  At the phenomenological level they are related to the inflaton
potential, although eventually one imagines there will be a theory with
specific fields and couplings, rather than simple mathematical forms.
Whether there can be different values for these inflationary parameters
in different volumes (or universes for that matter) is a topic of debate.

\subsection{Additional parameters}
\label{sec:extras}

Since the SMC is still being developed, and does not come with any sense of
completeness, then the adding of extra ingredients is limited only by the
imaginations of theorists.  At the moment, there is no compelling evidence for
adding any additional degrees of freedom when fitting the SMC to data.
However, it is certainly fruitful to continue to try, since this is the
only way that missing ingredients will be uncovered.
Among the less crazy ideas for adding features are: a component of
isocurvature perturbations; a free parameter for the primordial helium
abundance; features in the initial power spectrum; additional phenomenological
reionization parameters; parameters describing the evolution of
$w$ in some basis; or perturbations in the Dark Energy itself.

\section{The SMPP vs the SMC}
\label{sec:vs}
Although it is hard to resist the temptation to compare the two Standard
Models, the SMPP and the SMC are clearly quite different.  Probably the most
basic difference is that the SMPP, although it is called a `model', is actually
a `theory' -- in fact one of the best tested and most successful theories ever
devised -- while the SMC is very much just a model.  The SMPP predicted the
existence of the W and Z bosons, as well as a host of other features which
have now been precisely measured.  The SMC is certainly less mature, but in
fact there are a number of confirmed predictions which can already be claimed.
This is in addition to generic attributes of the observable Universe,
which simply relate to the Big Bang or hierarchical structure formation
frameworks.  At
least 7 such predictions could be advanced as successes of the SMC, starting
with the emergence of the CMB acoustic peaks (a process which began in
1994 \cite{ScoWhi}, and was certainly definitive by 2000 \cite{Pierpaoli00}).
These tests of the SMC are listed in Table~\ref{tab:conf}.\footnote{The first
two are often regarded as part of the basis of the SMC, rather than being
confirmed predictions, but this is not really the case, since they came after
the SMC was already essentially in place.}

\begin{table}
\captionwidth317pt
\begin{center}
\topcaption{Confirmed Predictions of the Standard Model of Cosmology.}
\label{tab:conf}
\begin{tabular}{lc}
\hline*
Prediction & Year of Confirmation \\
\hline*
CMB acoustic peaks & 1994 \\
Acceleration from SNe & 1998 \\
Cosmic shear & 2000 \\
CMB polarization & 2001 \\
High-$z$ cosmic jerk in SNe & 2001 \\
ISW-LSS correlations & 2003 \\
Baryon oscillations & 2005 \\
\hline*
\end{tabular}
\end{center}
\tablefootnote{Dates are for the first clear indication of the predicted
effect; in most cases much more significant confirmations soon followed.}
\end{table}

The two Standard Models are certainly distinct, since {\it none\/} of the
parameters of the SMC can be derived using the SMPP -- for example,
one needs to know at
least the value of $H_0$ (and then the full SMPP would give $\Omega_\nu$) or
the addition of $T_0$ and information about baryogenesis
(to get $\Omega_{\rm B}$).  It is also the case that very few of the
SMPP parameters have much bearing on the SMC, since only the physics of
photons, protons, neutrons and electrons is of primary importance for
physical cosmology, with
the neutrino sector being less important, and the rest being largely
irrelevant (except indirectly, through the relationship
with higher energy physics {\it beyond\/} the SMPP).

Because the sets of parameters are essentially disjoint, then
any fundamental answer to the question of `Life, the Universe
and Everything' will have to explain the parameters of {\it both\/}
Standard Models.

Another basic difference between the two Standard Models comes from how
they were developed.
The SMPP may be an incomplete theory (in the sense that one would like to
solve the `hierarchy problems', unify with gravity and include the extra
physics implied by the SMC), but at least it has a well-defined
mathematical structure.  The SMC, on the other hand is not something that
one might have dreamed up without peeking at the Universe.
Words like `unsatisfactory', `baroque' and `preposterous' are used often
invoked to describe the nature of the SMC.  However, as we have seen, the
SMC has to be taken quite seriously, since it has made several
testable predictions which have subsequently been confirmed, thus
conforming broadly to the scientific method.
While the more fundamental theory of high energy physics is
is expected to still contain the SMPP as its lower energy limit, the same
may not be true for the SMC within the ultimate Cosmological Model.
But given the SMC's decade of success, it is hard to see at this point
how the entire edifice can be askew.

Another apparent difference is
the sense in which the parameters of the SMPP are more
fundamental than those of the SMC,
although the distinction is perhaps becoming less clear.  This is connected
with the traditional view that in physics everything is exact, while in
astronomy everything is approximate.

In high energy physics one measures cross-sections,
decay rates, branching ratios, etc.~and relates these to the parameters of
the SMPP.  Every calculation is imagined to be correct,
up to a next-order term which has been neglected.  In
practice things are often more complicated, with some calculations seeming
to require non-perturbative approaches, but still, there is an implicit
assumption that there {\it are\/} exact values waiting to be measured.

In astronomy generally, the philosophy is quite different.  Instead of
measuring the properties of sub-atomic particles imbued with a few quantum
numbers, one is trying to measure the properties of macroscopic objects
over a huge range of mass scales, and hence approximate answers are often
sufficient (and intrinsic scatter in properties is expected).
But physical cosmology has become more precise than most other branches of
astrophysics, since one is attempting to get at a description of
the entire observable Universe by
measuring astrophysical quantities, which are related in a complicated way
to the underlying parameters of the SMC.

One accepts that there is a general level of uncertainty here, sometimes
referred to as `cosmic variance', meaning that the properties within our
Hubble volume may be slightly different from the expectation value averaged
over a large number of such volumes.  This idea is not too far from those
of the `Landscape', and hence it may be that practitioners of sub-atomic
physics may also have to get used to the idea of `cosmic variance'.

\section{Three Levels of Stochasticity}
\label{sec:stoch}

Discussion of the origin of the SMC parameters may help illuminate how
parameters are fixed in general.  One can imagine that some parameters
are entirely deterministic, i.e.~they are fixed by the mathematical
structure of the fundamental theory.  Other parameters may be more `accidental'
in nature, and it may be that if we can figure out where the
stochasticity comes in, that may help distinguish among theories
(see e.g.~\cite{Hogan00}).  But what does `stochastic' mean in the
context of the properties of the Universe?

\begin{table}
\captionwidth317pt
\begin{center}
\topcaption{The 3 Levels of Stochasticity.}
\label{tab:stoch}
\begin{tabular}{ll}
\hline*
Level & Affected quantities \\
\hline*
Observer epoch & $T_0$, $H_0$, $\Omega$s, \ldots \\
Cosmic variance & Dipole, other $a_{\ell m}$s, local $H_0$, $\Omega_{\rm tot}$,
 \ldots \\
Different vacua & $\Lambda$, inflation, SMPP, \ldots \\
\hline*
\end{tabular}
\end{center}
%\tablefootnote{}
\end{table}

There are 3 different levels of randomness that enter into this discussion
(see Table~\ref{tab:stoch}).
At the weakest level, some SMC parameters are stochastic in the sense that we
happen to measure them at a particular epoch.  There is no obvious property
which would have prevented us from asking about the Universe if we lived
in a civilisation several billion years into the past or the future.  If we
define some parameters to be simply the quantities at the time we
measure them\footnote{A fairly natural choice, at least for our present
purpose.}, then their values will be
partly accidental in the sense that the epoch we are living at contains some
randomness.  This may seem like a fairly ineffectual kind of stochasticity,
but it has the virtue of being easy to understand, and undeniably real.

A stronger level of stochasticity comes from the `cosmic variance' idea.
There is a certain volume in which we can make measurements, defined by the
region within which light can have reached us in the age of the Universe.
We {\it know\/} that each Hubble patch has an overall density fluctuation
of order $10^{-5}$, since the power spectrum gives roughly that amplitude
at `Horizon crossing'.  The density within our patch could easily be a
several $\sigma$ excursion from the average value, without it seeming terribly
unlikely.  This means that the global value of $\Omega_{\rm tot}$ can never be
determined to better than 4 decimal places.  And the same applies to all the
other $\Omega$s.  A more extreme example of this same principle
comes from the cosmic dipole -- one
can measure its value, but since its relative variance is large between
observers, then the value we measure is almost entirely a random number.
In general, {\it all\/} of the detailed 3-dimensional structure of the
observable Universe falls into this category.  Some initial fields chose
particular realisations consistent with some expectation values, and this
ultimately led to objects such as the Great Wall, the Local Supercluster,
the Milky Way galaxy, and the detailed configuration of matter within them.

The final level of randomness comes when one considers that there may be
multiple vacuum states into which different volumes fell.  Each of these
volumes would then have inflated in a different way and may have had different
values of the parameters of the SMPP.  Whether we consider these as separate
`universes' or as different volumes within a single space-time may be simply
semantics.

A useful outcome of considering the origin of the parameters of the SMC
may be that one is naturally led to the idea that several of the parameter
values contain a non-deterministic element, and that there are
different levels of randomness.  It may then be more reasonable to imagine
that quantities such as the vacuum energy could be stochastic in nature.
And so more generally,
the idea of different volumes having different SMPP parameters may seem like
less of a conceptual leap.  Seen in this light the details of the SMPP may
just be extensions of the SMP, as observed in our particular universe.

Whether the probability distributions of any quantities can be calculated
(or even defined) is altogether a different question.  A great deal has already
been said, and debated, about the reality of the Multiverse\footnote{A term
which certainly grabbed the imagination of science fiction writers, probably
even being used by them before it was adopted by physicists \cite{Moorcock}.},
and principles of selection, mediocrity, fine tuning, biophilia,
anthropy, Copernicanism and the like
\cite{Carter74,CarrRees79,Vilenkin95,Tegmark03,Kallosh03,Davies04,Hartle04,Smolin04,Aguirre05,Garriga05},
and little can be added here.

The main point to make is that our current
picture of the Universe contains elements of happenstance, and that
learning more about the cosmological model may help us to understand this
randomness, and hence the bigger picture.\footnote{Or then again, it may not.}
An extension of the idea
that stochasticity plays some role in establishing parameter values is
that we may be able to lump the parameters of the SMC and the SMPP together
as part of the set of observables of our corner of the multiverse.

\section{The Future}
\label{sec:future}

There are many ongoing and planned surveys which will improve the determination
of the parameters of the SMC.  These include new CMB experiments,
particularly the next generation satellite mission {\sl Planck\/}, scheduled
for launch in about 2 years.  The capabilities of {\sl Planck\/} mean that
it may be the experiment which takes us beyond the SMC.
There are several new experiments designed to
measure CMB polarization from either the ground or balloons, and there are
also ambitious plans for another satellite mission in the more distant
future to try to detect the inflationary B-modes.  Additionally,
there are aggressive plans for
measuring weak lensing, supernova distances and galaxy cluster abundances,
all of which should particularly help improve the uncertainties on $w$.

If the SMC is already in good shape, then why should we bother placing more
stringent bounds on its parameters?
The main reason to determine these quantities more precisely is that these are
{\it the\/} numbers that describe {\it the whole Universe},
and so they should just be measured better simply because they're
there.\footnote{And the Universe surely deserves at least as much
attention paid to its statistics as the average sports team.}
The secondary reason is that one
can only uncover inadequacies in the framework by making more precise
measurements.  Since the SMC is not self-contained in any way,
then there may well be major features which are currently missing.

Certainly there are many mysteries suggested by the SMC.  The obvious ones
are: What are the Dark Matter and Dark Energy?
How did baryogenesis work?
Did inflation really happen?

The first 3 of these issues may have resolutions at energy scales which
{\it might\/} be accessible at accelerators.  But the inflationary parameters
are probably probing energies ${\sim}\,10^{16}$GeV, far beyond the reach of
any terrestrial experiment.

The quantum fluctuations of the inflaton field generically have the right
properties to be the initial conditions for the CMB anisotropies, and indeed
to make all structures in the Universe today through gravitational
instability.  Inflation also solves some conceptual problems with
the Big Bang framework\footnote{Although admittedly
most people ignored them until there was a proposed solution!}, including
answering the `why is the Universe expanding today' problem -- the old answer
used to be `because it was expanding yesterday', but now there is the slightly
more satisfactory answer `because inflation started it off'.  However, there
is still the question of {\it why\/} the Universe inflated.  Or more
succinctly, `where did everything come from?'

Then there are the coincidence problems: Why is the Dark Energy starting
to dominate now?  And why is
$\Omega_{\rm CDM}\simeq5\Omega_{\rm B}$?  The first of these questions suggests
some selection in time, if not among vacuum states.  The second may be a clue
about physics beyond the SMPP, but whether it needs to have a stochastic
element is unclear.  Of course some coincidences
may not need an underlying explanation, e.g.~why is the density in stars
$\Omega_\star\simeq\Omega_\nu$ and why is $H_0 t_0$ so close to 1?
But other apparent coincidences exist where it is not obvious if an
explanation is required or not, e.g.~why is the redshift of matter-radiation
equality so close to that of recombination of the hydrogen plasma?

As well as tackling these `why'-type issues,
astrophysical cosmologists are busy trying to understand
how galaxies form and evolve, when and how the very first stars formed, and
what the relationship is between the formation of galaxies and their central
black holes.  This may seem like an entirely different endeavour from
determining the parameters of the SMC, but in fact it cannot be done in
isolation, since one needs to understand details of the non-linear tracers,
foregrounds and astrophysical processing of all cosmological signals in order
to measure the SMC parameters.

Given that the SMC raises more questions than it resolves, it seems
clear that there must be a bigger picture which remains to be uncovered.
It is truly surprising that the SMC has remained essentially unchanged for a
whole decade, but it seems
inconceivable that our current model is the final word.  It is, of course,
possible that some of today's list of 12 parameters will be found to be fixed
at their default values, at the same time that new degrees of freedom
need to be added in order to fit future data-sets.  There is no way to
predict how the final Cosmological Model will be different from the current
SMC.

With the rich pickings of cosmological data out there, the SMC will continue
to be probed with increasing precision.  One can see that the next decade
looks every bit as exciting as the last.  Eventually one can imagine a day
when we know precisely how many parameters are required and have a better
idea of how they are determined (including the level of randomness).

One can also take comfort in the fact that if we lived substantially
earlier or later in the history of the Universe it would be considerably
more difficult to make precise cosmological measurements from which to
quantify the Universe in which we live.  We do indeed live in interesting
times.


\begin{thebibliography}{9}
\bibitem{Aguirre05} A. Aguirre, M. Tegmark, J. Cosm. Astropart. Phys.,
 {\bf 0501}, 003 (2005).
\bibitem{AlpHer} R.A. Alpher, R.C. Herman, Phys. Today, {\bf 41}, No.$\,8$,
 p.$\,24$ (1988).
\bibitem{Astier05} P. Astier, et al., Astron. Astrophys., in press,
 astro-ph/0510447.
\bibitem{BarLoe} R. Barkana, A. Loeb, Phys. Rept., {\bf 349}, 125 (2001).
\bibitem{BarTip} J.D. Barrow, F.J. Tipler, The Anthropic Cosmological
 Principle, Clarendon Press, Oxford (1986).
\bibitem{BFPR} G.R. Blumenthal, S.M. Faber, J.R. Primack, M.J. Rees, Nature,
 {\bf 311}, 517 (1984).
\bibitem{Carr} B.J. Carr, Universe or Multiverse, Cambridge University Press
 (2005).
\bibitem{CarrRees79} B.J. Carr, M.J. Rees, Nature, {\bf 278}, 605 (1979).
\bibitem{Carroll01} S.M. Carroll, Living Rev. Rel., {\bf 4}, 1 (2001).
\bibitem{Carter74} B. Carter, in `Confrontation of Cosmological Theories
 with Observational Data', ed. M.S. Longair, Reidel, Dordrecht, p.$\,291$
 (1974).
\bibitem{Cole05} S. Cole, et al., Mon. Not. R. Astron. Soc., {\bf 362},
 505 (2005).
\bibitem{Davies04} P.C.W. Davies, Mod. Phys. Lett., {\bf A19}, 727 (2004).
\bibitem{Dickinson04} C. Dickinson, et al., Mon. Not. R. Astron. Soc.,
 {\bf 353}, 732 (2004).
\bibitem{EBW} G. Efstathiou, J.R. Bond, S.D.M. White, Mon. Not. R. Astron.
 Soc., {\bf 258}, P1 (1992).
\bibitem{Efstathiou90} G. Efstathiou, W.J. Sutherland, S.J. Maddox, Nature,
 {\bf 348}, 705 (1990).
\bibitem{PDG} S. Eidelman, et al., Phys. Lett. {\bf B592}, 1 (2004).
\bibitem{FieSar} B. Fields, S. Sarkar, in `The Review of Particle Physics',
 S. Eidelman, et al., Phys. Lett. {\bf B592}, 1 (2004).
\bibitem{Fixsen96} D.J. Fixsen, et al., Astrophys. J., {\bf 473}, 576 (1996).
\bibitem{Freedman01} W.L. Freedman, et al., Astrophys. J., {\bf 533}, 47 (2001).
\bibitem{Garriga05} J. Garriga, et al., hep-th/0509184.
\bibitem{Halverson02} N.W. Halverson, et al., Astrophys. J., {\bf 568}, 38
 (2002).
\bibitem{Hartle04} J.B. Hartle, in `The New Cosmology', ed. R.E. Allen, et al.,
 AIP Conf. Proc. Vol.~743, p.$\,298$ (2004) [gr-qc/0406104].
\bibitem{Hinshaw03} G. Hinshaw, et al., Astrophys. J. Suppl., {\bf 148}, 135
 (2003).
\bibitem{Hogan00} C.J. Hogan, Rev. Mod. Phys., {\bf 72}, 1149 (2000).
\bibitem{HuDod} W. Hu, S. Dodelson, Ann. Rev. Astron. Astrophys., {\bf 40},
 171 (2002).
\bibitem{HSSW} W. Hu, D. Scott, N. Sugiyama, M. White, Phys. Rev. {\bf D52},
 5498 (1995).
\bibitem{HuSugSil} W. Hu, N. Sugiyama, J. Silk, Nature, {\bf 386}, 37 (1997).
\bibitem{HuW} W. Hu, M. White, New Astron., {\bf 2}, 323 (1997).
\bibitem{Jones05} W. Jones, et al., Astrophys. J., in press, astro-ph/0507494.
\bibitem{Kallosh03} R. Kallosh, A. Linde, Phys. Rev. {\bf D67}, 023510 (2003).
\bibitem{Kofman93} L.A. Kofman, N.Y. Gnedin, N.A. Bahcall, Astrophys. J.,
 {\bf 413}, 1 (1993).
\bibitem{Kogut03} A. Kogut, et al., Astrophys. J. Suppl., {\bf 148}, 161
 (2003).
\bibitem{Kosowsky02} A. Kosowsky, M. Milosavljevic \& R. Jiminez, Phys. Rev.,
 {\bf D66}, 063007 (2002).
\bibitem{KraTur} L.M. Krauss, M.S. Turner, Gen. Rel. Grav., {\bf 27}, 1137
 (1995).
\bibitem{Kuo04} C.-L. Kuo, et al., Astrophys. J., {\bf 600}, 32 (2004).
\bibitem{Romans} C.R. Lawrence, D. Scott, M. White, Publ. Astron. Soc. Pac.,
 {\bf 111}, 525 (1999).
\bibitem{LytRio} D.H. Lyth, A. Riotto, Phys. Rept., {\bf 314}, 1 (1999).
\bibitem{MacTavish05} C.J. MacTavish, et al., Astrophys. J., in press,
 astro-ph/0507503.
\bibitem{Mather99} J.C. Mather, et al., Astrophys. J., {\bf 512}, 511 (1999).
\bibitem{Molaro02} P. Molaro, et al., Astron. Astrophys., {\bf 381}, L64 (2002).
\bibitem{Moorcock} M. Moorcock, `The Blood Red Game', Sci. Fict. Adv., Vol.~6,
 No.~32, p.$\,54$ (1963).
\bibitem{OstSte} J.P. Ostriker, P.J. Steinhardt, Nature, {\bf 377}, 600 (1995).
\bibitem{Peebles82} P.J.E. Peebles, Astrophys. J., {\bf 263}, L1 (1982).
\bibitem{Peebles84} P.J.E. Peebles, Astrophys. J., {\bf 284}, 439 (1984).
\bibitem{Peiris03} H.V. Peiris, et al., Astrophys. J. Suppl., {\bf 148}, 213
 (2003).
\bibitem{Pierpaoli00} E. Pierpaoli, D. Scott, M. White, Science, {\bf 287},
 2171 (2000) [astro-ph/0003393].
\bibitem{PisSer} O. Pisanti, P.D. Serpico, IFAE proceedings, in press,
 astro-ph/0507346.
\bibitem{Pope04} A.C. Pope, et al., Astrophys. J., {\bf 607}, 655 (2004).
\bibitem{Readhead04} A.C.S. Readhead, et al., Astrophys. J., {\bf 609}, 498
 (2004).
\bibitem{ReesCH} M.J. Rees, Our Cosmic Habitat, Princeton University Press
 (2001)
\bibitem{Rees04} M.J. Rees, in `Fred Hoyle's Universe', ed C. Wickramasinghe,
 et al., Kluwer, p.$\,95$ (2003) [astro-ph/0401424].
\bibitem{Riess04} A.G. Riess, et al., Astrophys. J., {\bf 607}, 665 (2004).
\bibitem{ScoSmo} D. Scott, G.F. Smoot, in `The Review of Particle Physics',
 S. Eidelman, et al., Phys. Lett. {\bf B592}, 1 (2004); in press (2006).
\bibitem{ScoWhi} D. Scott, M. White, in `2 years after {\sl COBE\/}', ed.
 L. Krauss, World Scientific, Singapore, p.$\,214$ (1994).
\bibitem{SSWZ} U. Seljak, N. Sugiyama, M. White, M. Zaldarriaga, Phys. Rev.
 {\bf D68}, 083507 (2003).
\bibitem{Seljak05} U. Seljak, et al., Phys. Rev. {\bf D71}, 103515 (2005).
\bibitem{Smolin04} L. Smolin, in `Universe or Multiverse', ed. B.J. Carr,
 Cambridge University Press (2005) [hep-th/0407213].
\bibitem{Smoot92} G.F. Smoot, et al., Astrophys. J., {\bf 396}, L1.
\bibitem{SmootScott02} G.F. Smoot, D. Scott, in `The Review of Particle
 Physics', K. Hagiwara, et al., Phys. Rev., {\bf D66}, 010001 (2002).
\bibitem{Spergel03} D.N. Spergel, et al., Astrophys. J. Suppl., {\bf 148},
 175 (2003).
\bibitem{Suss} L. Susskind, `The Anthropic Landscape of String Theory',
 hep-th/0302219.
\bibitem{Tegmark03} M. Tegmark, in `Science and Ultimate Reality: From Quantum
 to Cosmos', eds. J.D. Barrow, P.C.W. Davies \& C.L. Harper, Cambridge
 University Press (2003) [astro-ph/0302131].
\bibitem{LVWYM} L. van Waerbeke, Y. Mellier, Proceedings of the
 Aussois Winter School, in press, astro-ph/0305089.
\bibitem{Vilenkin95} A. Vilenkin, Phys. Rev. Lett., {\bf 74}, 846 (1995).
\bibitem{Vittorio85} N. Vittorio, J. Silk, Astrophys. J., {\bf 297}, L1 (1985).
\bibitem{WSS} M. White, D. Scott, J. Silk, Ann. Rev. Astron. Astrophys.,
 {\bf 32}, 319 (1994).
\bibitem{Resuscitated} M. White, D. Scott, J. Silk, M. Davis, Mon. Not., R.
 Astron. Soc., {\bf 276}, L69 (1995).
\bibitem{Zal03} M. Zaldarriaga, in `Measuring and Modeling the Universe',
 ed.~W.L. Freedman, Cambridge University Press, p.$\,310$ (2004)
 [astro-ph/0305272].
\end{thebibliography}
\end{document}